\documentclass[final,3p,times,twocolumn]{elsarticle}
\usepackage{amssymb}
\usepackage{color}
\usepackage{epstopdf}
\usepackage{graphicx}
\usepackage{dcolumn}
\usepackage{bm}
\usepackage{amsmath}
\usepackage{longtable}
\usepackage{ulem}
\usepackage{epstopdf}

\journal{Physics Letter B}

\begin{document}

\begin{frontmatter}

\title{Is Seniority a Partial Dynamic Symmetry in the First  $\nu g_{9/2}$ Shell?}

\author[1,2,3]{A.I.~Morales~\corref{cor1}}
\ead{Ana.Morales@ific.uv.es. }
\cortext[cor1]{Corresponding author.}

\author[2]{G.~Benzoni}
\author[4,5]{H.~Watanabe}
\author[6]{G.~de Angelis}
\author[5]{S.~Nishimura}
\author[7]{L.~Coraggio}
\author[7]{A.~Gargano}
\author[7,8]{N.~Itaco}
\author[9,10,11,12]{T.~Otsuka}
\author[9]{Y.~Tsunoda}
\author[13]{P.~Van Isacker}
\author[14,5]{F.~Browne}
\author[15]{R.~Daido}
\author[5]{P.~Doornenbal}
\author[15]{Y.~Fang}
\author[5]{G.~Lorusso}
\author[16,5]{Z.~Patel}
\author[16,5]{S.~Rice}
\author[17,5]{L.~Sinclair}
\author[5]{P.-A.~S\"oderstr\"om}
\author[18]{T.~Sumikama}
\author[6]{J.J.~Valiente-Dob\'on}
\author[5]{J.~Wu}
\author[10,5]{Z.Y.~Xu}
\author[15]{A.~Yagi} 
\author[10]{R.~Yokoyama}
\author[5]{H.~Baba}
\author[1,2]{R.~Avigo}
\author[19]{F.L.~Bello Garrote}
\author[2]{N.~Blasi}
\author[1,2]{A.~Bracco}
\author[14]{A.M.~Bruce}
\author[1,2]{F.~Camera}
\author[1,2]{S.~Ceruti}
\author[1,2]{F.C.L.~Crespi}
\author[20]{M.-C.~Delattre}
\author[21]{Zs.~Dombradi}
\author[6]{A.~Gottardo}
\author[5]{T.~Isobe}
\author[22]{I.~Kojouharov}
\author[22]{N.~Kurz}
\author[21]{I.~Kuti}
\author[16]{S.~Lalkovski}
\author[10]{K.~Matsui}
\author[23]{B.~Melon}
\author[24,25]{D.~Mengoni}
\author[10]{T.~Miyazaki}
\author[6]{V.~Modamio-Hoybjor}
\author[10]{S.~Momiyama}
\author[6]{D.R.~Napoli}
\author[10]{M.~Niikura}
\author[12,26]{R.~Orlandi}
\author[16]{Zs.~Podoly\'ak}
\author[16,27]{P.H.~Regan}
\author[5,10]{H.~Sakurai}
\author[19]{E.~Sahin}
\author[21]{D.~Sohler}
\author[22]{H.~Schaffner}
\author[10]{R.~Taniuchi}
\author[28,29]{J.~Taprogge}
\author[21]{Zs.~Vajta}
\author[2]{O.~Wieland}
\author[30]{M.~Yalcinkaya}

\address[1]{Dipartimento di Fisica dell'Universit\`{a} degli Studi di Milano, Via Celoria 16, 20133 Milano, Italy}
\address[2]{Istituto Nazionale di Fisica Nucleare, Sezione di Milano, Via Celoria 16, 20133 Milano, Italy}
\address[3]{IFIC, CSIC-Universitat de Val\`encia, E-46071 Val\`encia, Spain}
\address[4]{IRCNPC, School of Physics and Nuclear Energy Engineering, Beihang University, Beijing 100191, China}
\address[5]{RIKEN Nishina Center, 2-1 Hirosawa, Wako, Saitama 351-0198, Japan}
\address[6]{Istituto Nazionale di Fisica Nucleare, Laboratori Nazionali di Legnaro, I-35020 Legnaro, Italy}
\address[7]{Istituto Nazionale di Fisica Nucleare, Sezione di Napoli, Napoli, Italy}
\address[8]{Dipartimento di Matematica e  Fisica dell'Universit\`a degli Studi della Campania, I-81100 Caserta, Italy}
\address[9]{Center for Nuclear Study, The University of Tokyo, Bunkyo-ku, 113-0033 Tokyo, Japan}
\address[10]{Department of Physics, The University of Tokyo, Bunkyo-ku, 113-0033 Tokyo, Japan}
\address[11]{National Superconducting Cyclotron Laboratory, Michigan State University, East Lansing, MI 48824, USA}
\address[12]{Instituut voor Kern- en Stralingsfysica, Katholieke Universiteit Leuven, B-3001 Leuven, Belgium}
\address[13]{Grand Acc\'el\'erateur National d'Ions Lourds, CEA/DRF-CNRS/IN2P3, BP 55027, F-14076 Caen Cedex 5, France}
\address[14]{School of Computing, Engineering and Mathematics, University of Brighton, Brighton, United Kingdom}
\address[15]{Department of Physics, Osaka University, Osaka 560-0043 Toyonaka, Japan}
\address[16]{Department of Physics, University of Surrey, Guildford GU2 7XH, United Kingdom}
\address[17]{Department of Physics, University of York, Heslington, York YO10 5DD, United Kingdom}
\address[18]{Department of Physics, Tohoku University, Miyagi 980-8578, Japan}
\address[19]{Department of Physics, University of Oslo, N-0316 Oslo, Norway}
\address[20]{IPNO Orsay, 91400 Orsay, France}
\address[21]{MTA Atomki, H-4001 Debrecen, Hungary}
\address[22]{GSI, Planckstrasse 1, D-64291 Darmstadt, Germany}
\address[23]{INFN Sezione di Firenze, I-50019 Firenze, Italy}
\address[24]{Dipartimento di Fisica dell’Universit\`a degli Studi di Padova, I-35131 Padova, Italy}
\address[25]{Istituto Nazionale di Fisica Nucleare, Sezione di Padova, I-35131 Padova, Italy}
\address[26]{Advanced Science Research Center, JAEA, Tokai, Ibaraki 319-1195, Japan}
\address[27]{Nuclear Metrology Group, National Physical Laboratory, Teddington, Middlesex, TW11 0LW, UK}
\address[28]{Instituto de Estructura de la Materia, CSIC, E-28006 Madrid, Spain}
\address[29]{Departamento de F\'isica te\'orica, Universidad Aut\'onoma de Madrid, E-28049 Madrid, Spain}
\address[30]{Department of Physics, Istanbul University, 34134 Istanbul, Turkey}

\begin{abstract}
The low-lying structures of the midshell $\nu g_{9/2}$ Ni isotopes $^{72}$Ni and $^{74}$Ni have been investigated at the RIBF facility in RIKEN within the EURICA collaboration. Previously unobserved low-lying states were accessed for the first time following $\beta$ decay of the mother nuclei $^{72}$Co and $^{74}$Co. As a result, we provide a complete picture in terms of the seniority scheme up to the first $(8^+)$ levels for both nuclei. The experimental results are compared to shell-model calculations in order to define to what extent the seniority quantum number is preserved in the first neutron $g_{9/2}$ shell. We find that the disappearance of the seniority isomerism in the $(8^+_1)$ states can be explained by a lowering of the seniority-four $(6^+)$ levels as predicted years ago. For $^{74}$Ni, the internal de-excitation pattern of the newly observed $(6^+_2)$ state supports a restoration of the normal seniority ordering up to spin $J=4$. This property, unexplained by the shell-model calculations, is in agreement with a dominance of the single-particle spherical regime near $^{78}$Ni. 
\end{abstract}


\end{frontmatter}


The pairing interaction, first introduced by Racah for the classification of $n$ electrons in atoms~\cite{Rac43}, can be treated as a first-order approximation of the strong residual force between identical nucleons \cite{Rac52,Flo52}. The so-called \textsl{nuclear pairing} provides the grounds for one of the simplest and most powerful approaches to the nuclear shell model, the seniority scheme \cite{Talmi,Casten}. The seniority $\upsilon$ refers to the number of protons or neutrons that are not in pairs with total angular momentum $J=0$. Using the seniority approach, the $jj$ coupling of nucleons in a single $j$ shell can be classified by two quantum numbers, the total spin $J$ and the seniority $\upsilon$. 

The conservation of seniority results in insightful properties that can be used to identify shell closures far from stability \cite{Res04,Esc06}. The best known is that the $\upsilon=2$ excitation energy spectrum in a single $j$ shell is constant, no matter what the number of valence particles is. An equally important property is that the quadrupole operator between states of seniority $\upsilon=2$ vanish in midshell, leading to the appearance of seniority isomers arising from the fully-aligned configuration of one neutron or proton pair  \cite{Gra02,Gor97,Grz98,Maz05,Jun07,Got12,Wat13,Sim14}. 

Seniority is a good quantum number for any two-body interaction in orbitals with $j\leqslant7/2$ \cite{Talmi,Casten}. The first shell in which it might not be conserved is $g_{9/2}$ \cite{Esc06,Qi11}, although it is still valid for a subset of solvable eigenstates \cite{Esc06,Qi11,Zam07,Isa08,Qi10,Isa13}. This property can be seen as an example of a partial dynamic symmetry \cite{Alh92,Lev96}. In particular, for a four-particle (hole) system, two special states with good seniority $\upsilon=4$ and total spins $J=4$ and $6$ are found as eigenstates of any two-body interaction \cite{Isa14}.   

Excellent test cases for the partial conservation of seniority are the semi-magic Ni isotopes with four neutrons ($^{72}$Ni) and four neutron holes ($^{74}$Ni) in the first $\nu g_{9/2}$ shell. While the $\upsilon=2$, $8^+$ seniority isomers in $^{70}$Ni and $^{76}$Ni have lifetimes consistent with the seniority scheme \cite{Grz98,Sod15}, their analogs in $^{72}$Ni and $^{74}$Ni were not observed in the expected $\mu s$ time range \cite{Maz05,Saw03}. This astonishing deviation was attributed to a quasi-degeneracy of the $6^+$ level with seniority $\upsilon=2$ and the special $6^+$ state with seniority $\upsilon=4$, the energy lowering of the latter being ascribed to an increased contribution of proton excitations across the $Z=28$ gap \cite{Gra02,Isa13,Lis04}. The recent measurement of key experimental observables such as the energy of the yrast $8^+$ state and the lifetime of the $4^+_1$ level in $^{72}$Ni \cite{Mor16,Kol16} supports this explanation. 
In spite of the great advances, the question regarding the conservation of seniority as a good quantum number for close-lying levels with equal total spin $J$ is still under debate \cite{Mac17,Qi17}. 

In order to give insight into this issue,  we have investigated the $\beta$ decays of $^{72}$Co and $^{74}$Co at the RIBF facility operated by the RIKEN Nishina Center and the Center for Nuclear Study of the University of Tokyo in Japan. We report on a wealth of new spectroscopic information in the daughter nuclei $^{72}$Ni and $ ^{74}$Ni. For $^{74}$Ni only two $\gamma$ rays attributed to the decay of the $(2^+_1)$ and $(4^+_1)$ states were previously known \cite{Maz05,Mar14}, while for $^{72}$Ni a comprehensive decay level scheme was reported previously by some of the authors \cite{Mor16}. Here we report for the first time the half-life of the $(6^+_1)$ state, which was measured using in-flight $\beta$-delayed fast-timing spectroscopy \cite{Bro15,Bro17}. 
The present  experimental results bring  to completion the  knowledge gained  by the EURICA collaboration on the decay of neutron-rich odd-odd Co isotopes \cite{Sod15,Mor16,Mor17}.

The soundness of the seniority scheme up to the $J^{\pi}=(8^+_1)$ states in $^{72}$Ni and $^{74}$Ni will be discussed in the context of four shell-model calculations \cite{Isa14,Lis04,Tsu14,Cor14} that have recently been used to describe the properties of nuclei in the region \cite{Mor16,Kol16,Mar14,Mor17,Leo17,Sah17,Ben15}. The first is the Monte-Carlo Shell Model (henceforth called MCSM) \cite{Tsu14,Ots16} based on the A3DA Hamiltonian \cite{Shi12} including the $fpg_{9/2}d_{5/2}$ model space for both protons and neutrons. The second is based on a phenomenological effective interaction derived for the $f_{5/2}pg_{9/2}$ neutron model space (referred to as SM Lis) \cite{Lis04}. The third (called SM 2+5) is a realistic shell-model calculation performed with a two-body effective interaction derived within the  framework of many-body perturbative theory starting from the CD-Bonn NN potential \cite{Cor09,Cor13} in a valence space consisting of  the $f_{7/2}$ and $p_{3/2}$ orbitals for protons and the $fpg_{9/2}d_{5/2}$ orbitals for neutrons. The last, called SM SCI, is based on an empirical interaction in the $(\nu g_{9/2})^n$ shell, with sliding two-body matrix elements that interpolate between $^{70}$Ni and $^{76}$Ni \cite{Isa14}.

\begin{figure*}
\centering
\includegraphics[width=\textwidth]{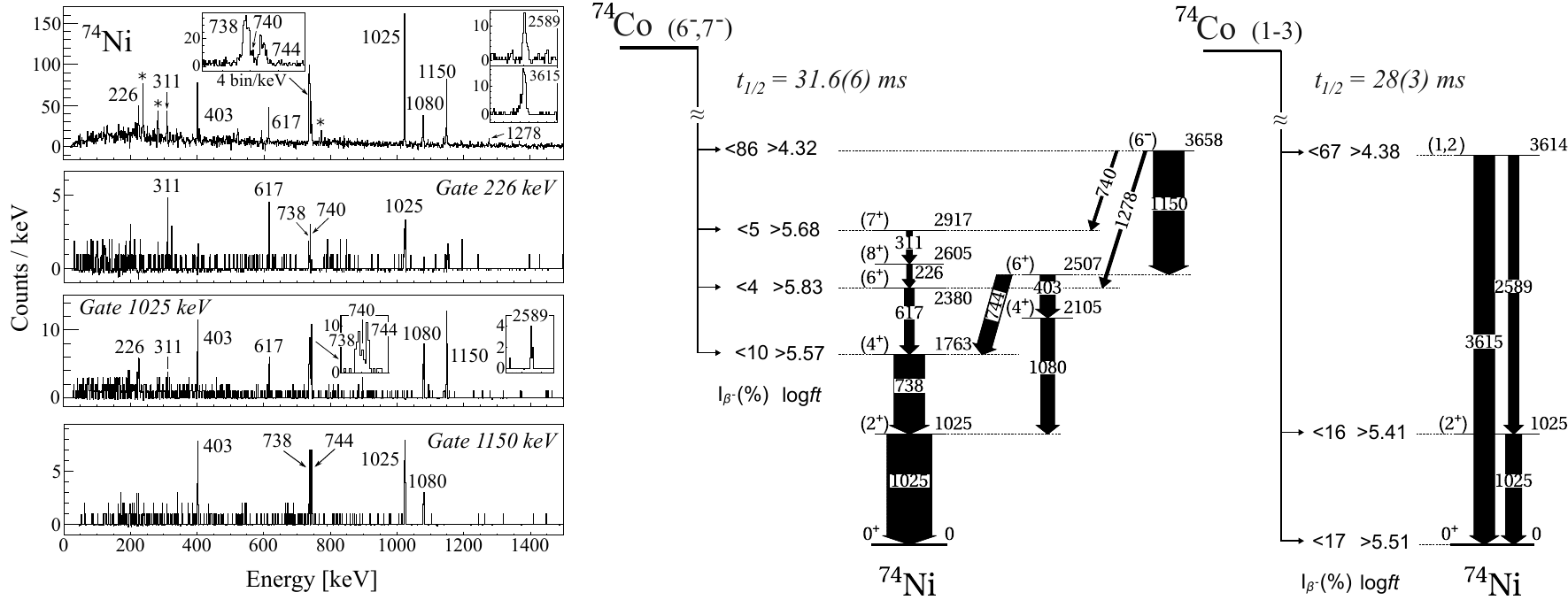}   
\caption{(Left) From top to bottom, $\beta$-delayed $\gamma$-ray energy spectrum following the implantation of $^{74}$Co during a time interval of 170 ms and $\gamma$-$\gamma$ coincidence spectra gated on the 226-, 1025-, and 1150-keV $\gamma$ transitions of $^{74}$Ni. The asterisks indicate $\gamma$ rays following $\beta$-delayed neutron emission. (Center) Partial level scheme of $^{74}$Ni attributed to the $\beta$ decay of the high-spin and (Right) low-spin isomer of $^{74}$Co. Tentative spin and parities $J^{\pi}$ of the observed states are indicated on the left part of the levels. Widths of the arrows are proportional to their absolute intensities (normalized to the number of nuclei produced in the high- and low-spin states, respectively). In both cases the log$ft$ values are calculated assuming that the $\beta$-decaying state is the ground state in $^{74}$Co. The $Q_{\beta}=15640(540)$ keV is taken from Ref. \cite{Wan17}.}
\label{fig1}
\end{figure*}

A $^{238}$U beam was accelerated up to an energy of 345~MeV/nucleon by the RIBF cyclotron accelerator complex before impinging on a 3-mm beryllium target. Radioactive secondary beams were produced by in-flight fission. The primary beam intensity was 10~pnA at the production target. Fission products were separated based on their momenta and mass-to-charge ratios (A/Q) in the BigRIPS separator using a 6-mm-thick Al achromatic degrader placed at the dispersive focal plane. Particle identification was achieved by measuring the energy loss, magnetic rigidity, and time-of-flight of the transmitted fission residues which provided $Z$ and $A/Q$ information on an event-by-event basis~\cite{Fuk13}.

The identified nuclei were implanted in the WAS3ABi active stopper after traversing the ZeroDegree spectrometer \cite{ZeroDegree}. WAS3ABi consisted of a compact stack of 5 double-sided Si strip detectors (DSSSD) divided into 60 vertical (X) and 40 horizontal (Y) strips of 1-mm pitch and 1-mm thickness each ~\cite{Nis12}. At a distance of 22 cm, the EURICA $\gamma$-ray detector array~\cite{Sod13} surrounded the active stopper. It was composed of 12 HPGe cluster detectors with 7 crystals each, providing an absolute detection efficiency of about 11$\%$ at 662 keV. The setup was implemented with a fast-timing array of 18 LaBr$_3$(Ce) detectors for the identification of $\gamma$ rays and two fast-plastic scintillators for the detection of $\beta$ particles. The latter were placed at 2-5 mm from the first and last DSSSD detectors, respectively~\cite{Mor15_proc}.

Implanted nuclei were identified in WAS3ABi as overflow energy signals in both X and Y strips, while $\beta$ signals were read using high-gain analog electronics. Their time and position (DSSSD, X and Y strips) were recorded event-by-event in order to build spatial and time correlations between implanted nuclei and $\beta$ particles. The time and energy of $\gamma$ rays from decay successors were registered by EURICA. 
Nuclear half-lives were obtained from the time differences between $\beta$ particles and de-exciting $\gamma$ rays, i.e., the moments of formation and de-excitation of the states, respectively. The $\beta$-$\gamma$ time differences obtained with the fast-timing array were corrected for standard time-walk effects and other dependences related to the use of the $\beta$ electron as the reference start signal, namely its energy and position. The final time resolution of the system was 300 ps (FWHM). This allowed for the measurement of half-lives down to about 100 ps \cite{Pat14}. Almost $9\times10^5$ and $2\times10^4$ implantation events were registered for $^{72}$Co and $^{74}$Co, respectively. 

The $\gamma$-ray energy spectrum following the $\beta$ decay of $^{74}$Co to $^{74}$Ni is shown in the top left panel of Fig. \ref{fig1}. Sorting conditions included a maximum ion-$\beta$ time difference of 170~ms and a maximum distance of (X$\pm$1,Y$\pm1$) strips.  The implantation depth was also constrained to the same DSSSD. For the $\gamma$-$\gamma$ coincidence analysis, the maximum time difference between $\gamma$ rays was set to 300 ns. Coincidence spectra gated on key transitions, namely the proposed $(8^+_1)\rightarrow(6^+_1)$, $(2^+_1)\rightarrow(0^+_1)$, and $(6^-_1)\rightarrow(6^+_2)$ $\gamma$ rays at 226, 1025, and 1150 keV, respectively, are shown in the three bottom panels on the left of Fig. \ref{fig1}. This information was used to build the level schemes following the $^{74}$Co$\rightarrow^{74}$Ni decay, which are shown in the central and right panels of Fig. \ref{fig1}. The proposed spin and parity assignments are based on the strong resemblance with the $\beta$ decay $^{72}$Co$\rightarrow^{72}$Ni \cite{Mor16}. One can see that the 403-, 617-, 738-, 744-, 1080-, and 1150-keV $\gamma$ rays de-excite the high-spin state at 3658 keV, which is strongly fed in the $\beta$ decay of $^{74}$Co. The 2589- and 3615-keV transitions de-excite the low-spin state at 3614 keV, also highly fed in the $\beta$ decay of $^{74}$Co. The internal ($\gamma$) de-excitation pattern of these two levels provides experimental evidence for the existence of two $\beta$-decaying states at high and low spins in $^{74}$Co. A $\gamma$-ray time-behavior study results in comparable $\beta$ half-lives,   $t_{1/2}=31.6(6)$ ms for the high-spin state and $t_{1/2}=28(3)$ ms for the low-spin one.  The $\beta$ and $\gamma$ intensities for each decay are shown in Tables I and II of the supplementary material.

\begin{figure}
\vspace*{0.1cm}
\hspace*{-0.3cm}
\centering
\includegraphics[width=7.7 cm]{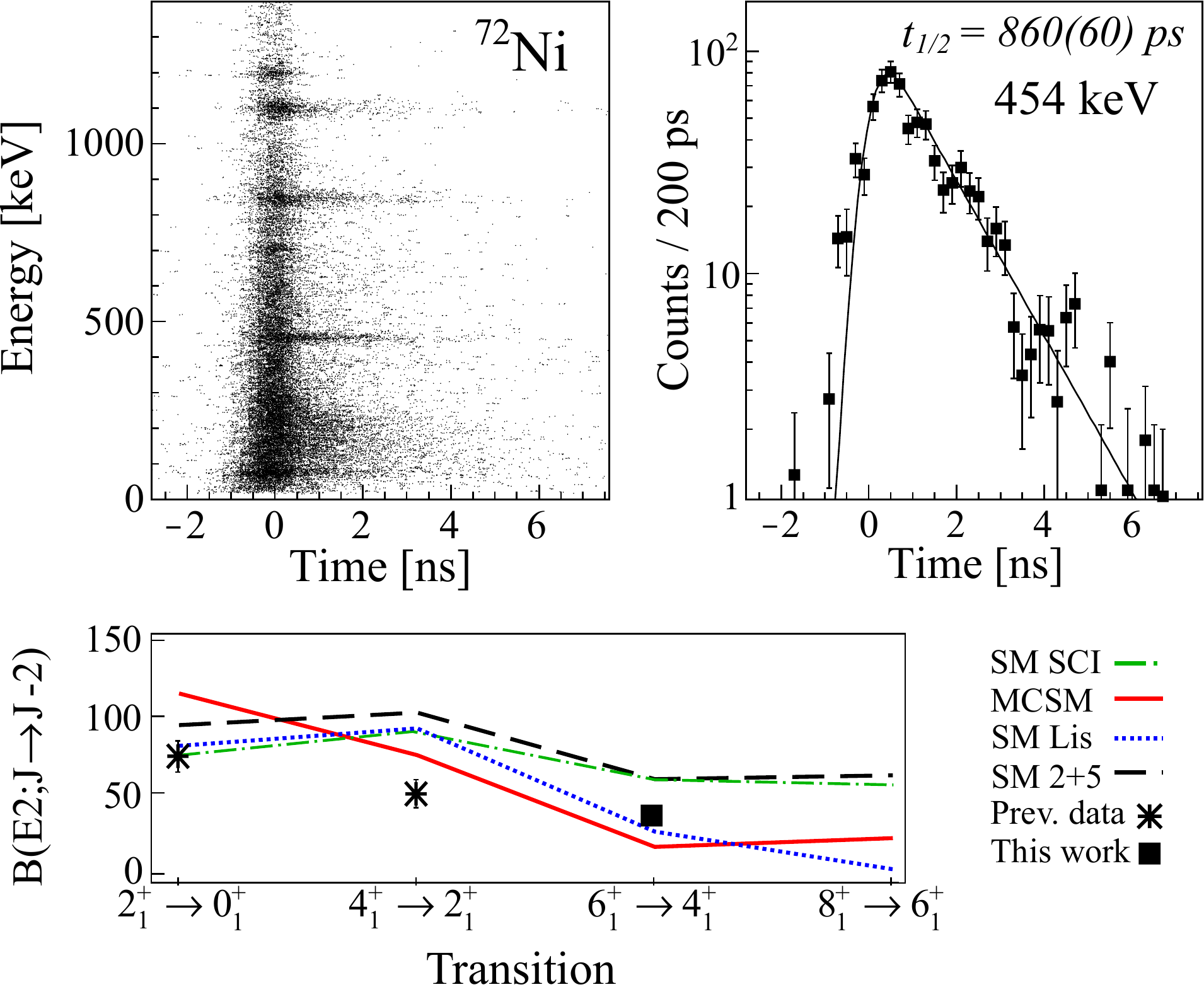}   
\caption{(Color on-line) Top left: Energy-time matrix showing delayed $\gamma$ transitions for the decay $^{72}$Co$\rightarrow^{72}$Ni. The $\gamma$-ray energy is plotted as a function of the decay time of the transition. Top Right: Time projection of the energy-time matrix gated on the 454-keV $\gamma$ ray. Close-lying Compton background has been subtracted as indicated in Ref. \cite{Mor17b}. The least-squares fit to a convoluted Gaussian plus exponential function is also shown. Bottom: Comparison of experimental and theoretical B(E2)$\downarrow$ values (expressed in units of e$^2$fm$^4$). See text for details.}
\label{fig2}
\end{figure}

\begin{figure*}
\centering
\includegraphics[width=\textwidth]{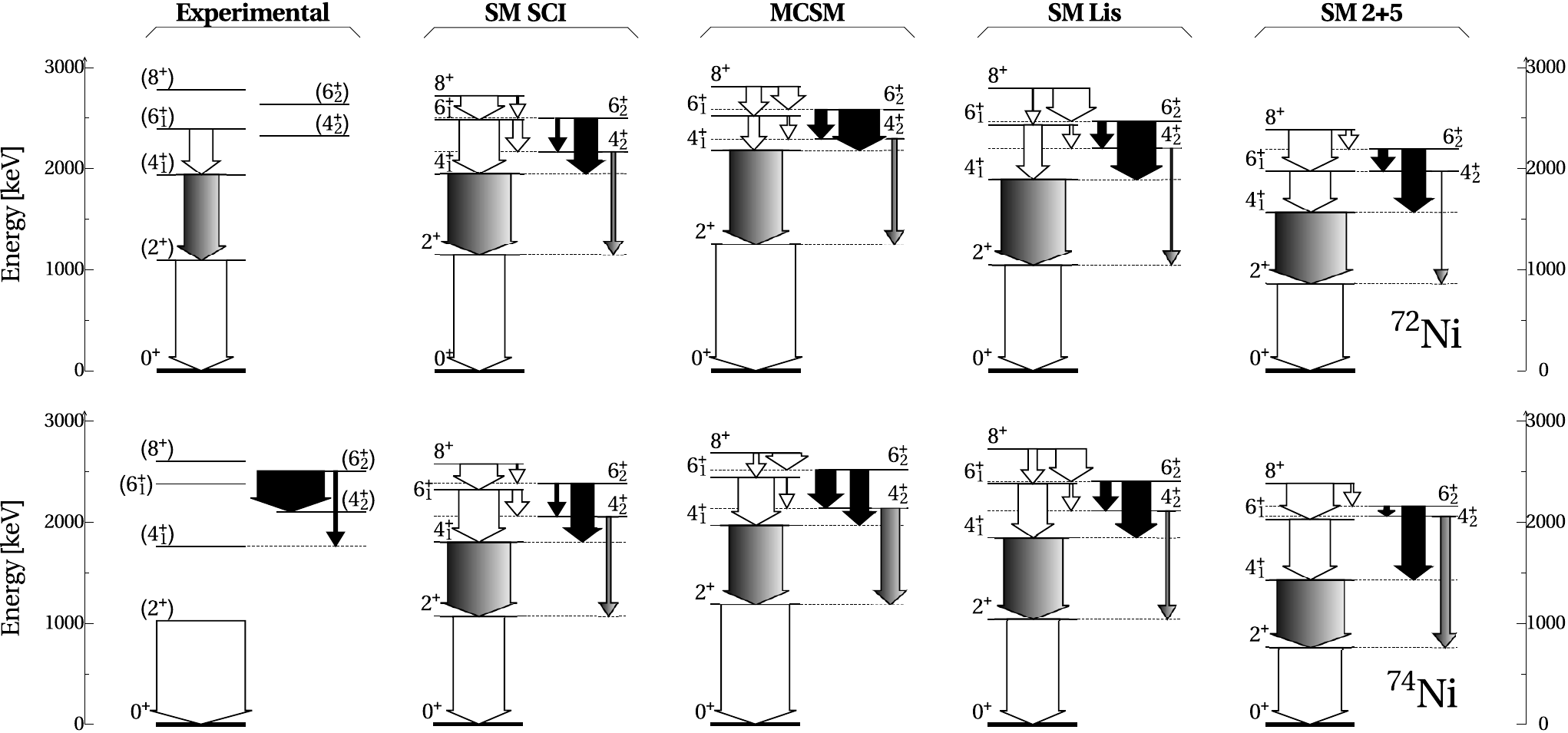}   
\caption{Experimental (\cite{Kol16,Mar14} and present results) and theoretical \cite{Isa13,Tsu14,Lis04,Cor14} level schemes in $^{72}$Ni (top) and $^{74}$Ni (bottom) centered in on the seniority states up to $J^{\pi}=8^+$. The thickness of the arrows is proportional to the B(E2) values of the transition connecting the states. The SM SCI predictions are scaled to the measured B(E2,$2^+_1\rightarrow0^+$) strength in $^{72}$Ni \cite{Kol16}. For the experimental  $(6^+_2)\rightarrow(4^+_1)$ and $(6^+_2)\rightarrow(4^+_2)$ decays in $^{74}$Ni, the B(E2) values are normalized to the B(E2;$6^+[\upsilon=2]\rightarrow 4^+[\upsilon=2])$ strength calculated by SM SCI, which is predicted to connect the $6^+_2$ and $4^+_2$ states by the calculation. Shaded and filled arrows indicate the B(E2) values relevant in eqs. \ref{eq1} and \ref{eq2}, respectively. See text for details.}
\label{fig3}
\end{figure*}

It is to note that in $^{74}$Co the $\nu g_{9/2}$ orbital  is well over half-filled (with three neutron holes) and the parabolic splitting of the $\pi f_{7/2}^{-1}\otimes \nu g_{9/2}$ multiplet may have inverted its orientation with respect to the proton hole\textendash neutron particle system \cite{Paa79,Van85}. Similarly as in $^{76}$Co \cite{Sod15}, the two  isomeric states reported here may correspond to the lowest-lying members of the multiplet, the $8^-$ and $1^-$ states. The $8^-$ level will mainly decay to the $7^-$ state built on the $(\nu f_{5/2}^{-1}\nu g_{9/2})$ configuration, thus converting a $f_{5/2}$ neutron into a $f_{7/2}$ proton via an allowed Gamow-Teller transition. The $7^-$ state is predicted at 4350 keV by the SM Lis calculation. Alternatively, the $8^-$ level  in $^{74}$Co can decay  through a $\nu g_{9/2}\rightarrow \pi f_{7/2}$ first-forbidden transition to the seniority-two $8^+_1$ state in $^{74}$Ni, which is predicted at 2728 keV. But, if the multiplet has not flipped yet, the ground state in $^{74}$Co will more likely have $J^{\pi}=6^-$ or $7^-$ as for the lighter Co isotopes \cite{Mol00,Mor16}, decaying to the $6^-$ member of the $(\nu f_{5/2}^{-1}\nu g_{9/2})$ configuration in $^{74}$Ni, which is predicted at 3859 keV by the calculation. 

The large feeding to the experimental level at 3658 keV (log$ft>4.32$, see Fig. \ref{fig1}) indicates the occurrence of an allowed Gamow-Teller transition. This, together with the fact that the state decays internally to the $(6^+)$ and $(7^+)$ candidates, supports a $(6^-)$ assignment for the 3658-keV level in $^{74}$Ni and hence a $(6^-)$ or $(7^-)$ nature for the mother state in $^{74}$Co. Further support comes from the SM 2+5 calculation, which predicts the $7^-$ level in $^{74}$Co to decay by $\beta$ emission. Indeed, only the $J^{\pi}=0^-$ ground state is expected below the $7^-$ $\beta$-decaying isomer located at 103 keV. Based on these arguments, we conclude that the multiplet has not flipped yet in $^{74}$Co.

The internal decay of the $(6^-)$ level has a branch to the yrast sequence $(8^+)\rightarrow(6^+)\rightarrow(4^+)\rightarrow(2^+)\rightarrow(0^+)$  through the $(7^+)$ state, as in $^{72}$Ni \cite{Mor16}. At variance with this case, the strongest gamma transition from the $(6^-)$ level feeds the $(6^+_2)$ state. Meanwhile, the low-spin state of $^{74}$Co decays to the 3614-keV level in $^{74}$Ni through a strong Gamow-Teller transition (log$ft>4.38$, see Fig. \ref{fig1}). The internal $\gamma$ de-excitation occurs either through the first $(2^+)$ level or directly to the ground state. This limits the spin of the 3614-keV level to $(1)$ or $(2)$ and that of the low-spin isomer of $^{74}$Co to $J^{\pi}\leq(3)$, which is consistent with the predictions of the SM 2+5 calculation.

For the decay $^{72}$Co$\rightarrow^ {72}$Ni, we have performed the first in-flight $\beta$-delayed fast-timing study. We have applied the same sorting conditions as for $^{74}$Ni, except for the maximum ion-$\beta$ correlation time, set to 270~ms \cite{Mor16}. The top-left panel of Fig. \ref{fig2} shows the $\gamma$-ray energy signals registered in the LaBr$_3$(Ce) detectors as a function of the $\beta$-$\gamma$ time differences measured with the fast-timing system for this decay. Three $\gamma$-ray transitions are unambiguously observed as delayed: The $(6^+_1)\rightarrow(4^+_1)$ at 454 keV, the $(4^+_1)\rightarrow(2^+_1)$ at 843 keV, and the  $(2^+_1)\rightarrow0^+_1$ at 1095 keV. We conclude, then, that the $(6^+_1)$ state in $^{72}$Ni is isomeric. A least-squares fit of the time spectrum gated on the 454-keV transition to a Gaussian plus exponential convolution function yields a half-life of $t_{1/2}^{6^+}=860(60)$~ps, see the top-right panel of Fig. \ref{fig2}. The corresponding reduced transition probability, B(E2$)=34(2)$~e$^2$fm$^4$, is indicated by a square in the bottom panel, while previously measured B(E2) values for the $(2^+_1)\rightarrow 0^+_1$ and $(4^+_1)\rightarrow(2^+_1)$ transitions \cite{Kol16} are shown in asterisks. For the sake of comparison, we also show the  B(E2) strengths calculated with the SM SCI (dash-dotted green line), MCSM (solid red line), SM Lis (dotted blue line), and SM 2+5 (dashed black line). The proton and neutron effective charges of the electric quadrupole operator E2 used in each shell-model calculation are indicated in Table III of the supplementary material. Note that for SM SCI we show B(E2) values normalized to the experimental B(E2$;2^+_1\rightarrow 0^+_1$) strength. 

In general, all the calculations qualitatively reproduce the systematics of the experimental results. The best quantitative agreement is found for SM Lis and MCSM, although only the latter follows the smooth downward trend of the measured B(E2$;J\rightarrow J-2$) values. This highlights the role of the proton core excitations in the description of the electromagnetic transition rates for the Ni isotopes.

It is to note that $^{72}$Ni and $^{74}$Ni show a different decay pattern for the $(6^+_2)$ state. While in $^{72}$Ni the main de-excitation branch feeds the $(4^+_1)$ state \cite{Mor16}, in $^{74}$Ni the $(4^+_1)$ and $(4^+_2)$ levels are fed with equal intensity, $I_{\gamma}^{403}=I_{\gamma}^{744}=36(6)\%$  (see Table I in the supplementary material). Although we cannot determine absolute B(E2) values for the $6^+_2\rightarrow4^+_2$ and $6^+_2\rightarrow4^+_1$ transitions in $^{74}$Ni, their ratio $R = \tfrac{B(E2,6^+_2\rightarrow4^+_2)}{B(E2,6^+_2\rightarrow4^+_1)}$ can be extracted from the experimental $\gamma$ intensities and the energies of the two $\gamma$ rays, resulting in $R = 22(5)$. 

In the following we will discuss the seniority of the $4^+$ and $6^+$ states in $^{72}$Ni and $^{74}$Ni by comparing the experimental and calculated B(E2) ratios, which are provided in tables IV and V of the supplementary material, with the ratios $R_{42}$ and $R_{64}$ obtained assuming exact conservation of seniority \cite{Isa08,Isa13,Isa14,Qi17}:

\begin{equation}
\label{eq1}
R_{42}= \dfrac{B(E2;4^+[\upsilon=4]\rightarrow 2^+[\upsilon=2])}{B(E2;4^+[\upsilon=2]\rightarrow 2^+[\upsilon=2])}\approx14
\end{equation}
\begin{equation}
\label{eq2}
R_{64}=\dfrac{B(E2;6^+[\upsilon=2]\rightarrow 4^+[\upsilon=4])}{B(E2;6^+[\upsilon=2]\rightarrow 4^+[\upsilon=2])}\approx7.6
\end{equation}

Here, we will make a qualitative discussion based on Fig. 3, where the experimental and theoretical states up to $J^{\pi}=8^+$ are shown for $^{72}$Ni (top) and $^{74}$Ni (bottom). In the figure, all the calculations predict the B(E2$;4^+_1\rightarrow 2^+$) value greater than the B(E2$;4^+_2\rightarrow 2^+$) one. Hence, according to eq. (\ref{eq1}), the calculated yrast $4^+$ states can be interpreted as the special $J^{\pi}=4^+$, $\upsilon=4$ levels that conserve seniority for any two-body interaction \cite{Isa14}. It is to note that the lowering of the $6^+$, $\upsilon=4$ states was attributed to the low energy of the $2^+$ level in previous studies \cite{Saw03}. The large value of the two-body matrix element $\nu g_{9/2}^2, J^{\pi}=2^+$ in the SM 2+5 calculation supports this interpretation, which can be extended also to the $4^+$, $\upsilon=4$ states.


In $^{74}$Ni, the experimental ratio $R=22(5)$ provides evidence for the hindrance of the $(6^+_2)\rightarrow (4^+_1)$ transition. This can only be explained within the seniority scheme if the corresponding E2 quadrupole operator connects levels of equal seniority. For the $(6^+_2)$ level, we assume main seniority $\upsilon=2$ based on the absence of isomerism in the $(8^+)$ state \cite{Maz05} and the non-observation of the competing $(8^+_1)\rightarrow (6^+_2)$ $\gamma$ branch (see Fig. \ref{fig1}). Hence, the most probable seniority for the $(4^+_1)$ level is  $\upsilon=2$, resulting in a recovery of the normal seniority ordering in the $(4^+)$ states that is further supported by the modest agreement with the $R_{64}\approx7.6$ ratio expected for a seniority-conserving interaction (see eq. \ref{eq2}). This result differs from the one found in $^{72}$Ni. In this case, evidence for the inversion of seniorities in both $(6^+)$ and $(4^+)$ states is reported in the literature \cite{Maz05,Mor16,Kol16}.

In $^{74}$Ni, the deviation of a factor $\sim$3 between the experimental result and the seniority-conserving prediction may be due either to configuration mixing  with close-lying orbitals or to excitations across the $Z=28$ and $N=50$ shell gaps. Indeed, the addition of two neutrons induces a mixing of neutron configurations that can modify the proton distribution with respect to $^{72}$Ni. The inclusion of core excitations in the calculations, thus, helps improving the description of the experimental ratio $R$ and eventually explain the evolution of the seniority ordering in the $4^+$ states of $^{72}$Ni and $^{74}$Ni. In table IV of the supplementary material one can see that the best, although still poor agreement with the experimental ratio $R$ is found for MCSM, the only calculation including the full proton $fpg_{9/2}d_{5/2}$ model space and the neutron $d_{5/2}$ shell above $N=50$. Similarly, the hindrance of the $4^+(\upsilon=2) \rightarrow 2^+(\upsilon=2)$ transition in the valence-mirror-symmetry nucleus $^{96}$Pd has been attributed to core excitations across $N=50$ \cite{Mac17}.

Last but not least, seniority distortions might be due to the presence of proton-neutron interactions that do not conserve seniority. In $^{74}$Ni, the MCSM calculation predicts a $\gamma$-soft deformed minimum stabilized by many particle-hole excitations across the $Z=28$ and $N=40$ shell gaps. This new type of shell evolution, known as ``type II'' \cite{Tsu14}, has recently been called for to explain the existence of a deep prolate minimum in nearby nuclei  \cite{Mor17,Leo17,spoiler,Cri16}. In $^{74}$Ni, the $0^+_2$ bandhead of the $\gamma$-soft shape is expected at an energy of 2301 keV, comparable to those of the $(4^+_2)$, $(6^+_1)$, and $(6^+_{2})$ levels at 2104, 2380, and 2507 keV, respectively. This proximity might favor the appearance of seniority non-conserving proton-neutron interactions that result in a larger hindrance of the $(6^+_2)\rightarrow (4^+_1)$ transition. 

Summarizing, we have observed for the first time the sought-after $8^+$ seniority-two states in the midshell $\nu g_{9/2}$ $^{72}$Ni and $^{74}$Ni isotopes. The disappearance of the seniority isomerism in these levels can be understood, as foreseen many years ago \cite{Gra02}, in terms of a strong reduction in the excitation energy of the special $(6^+)$ states with seniority $\upsilon=4$, which have a lower excitation energy than the $(6^+)$ states with seniority $\upsilon=2$. In $^{74}$Ni, evidence for the recovery of the normal seniority ordering up to spin $J=4$ is provided. This is in agreement with a smoother reduction of the $Z=28$ shell gap from $N=40$ to 50 \cite{Sah17}. Since the present experimental results cannot be reproduced by a variety of shell-model calculations employing different interactions and model spaces, further quantitative knowledge on the electromagnetic transition rates between the yrast and yrare $(6^+)$ and $(4^+)$ states is desirable. 


The excellent work of the RIKEN accelerator staff for providing a stable and high intensity $^{238}$U beam is acknowledged. We acknowledge the EUROBALL Owners Committee for the loan of germanium detectors and the PreSpec Collaboration for the readout electronics of the cluster detectors. Part of the WAS3ABi was supported by the Rare Isotope Science Project (RISP) of the Institute for Basic Science (IBS), funded by the Ministry of Science, ICT and Future Planning (MSIP) and the National Research Foundation (NRF) of South Korea (Grant No. 2013M7A1A1075764). The MCSM calculations were performed on the K computer at the RIKEN AICS (Project ID: hp150224). This work is co-financed by the European Union, the European Social Fund, and the European Regional Development Fund. Support from the Italian Ministero dell'Istruzione, dell'Universit\`a e della Ricerca through Programmi di Ricerca Scientifica di Rilevante Interesse Nazionale (PRIN) No. $2001024324\_01302$, the NuPNET-ERA-NET within the NuPNET GANAS project under grant agreement No. 202914, and the European Union within the 7th Framework Program FP7/2007-2013 under grant agreement No. 262010 ENSAR-INDESYS is acknowledged. This work was also financed by the Spanish Ministerio de Ciencia e Innovaci\'on under Contracts s No. IJCI-2014-19172, No. FPA2009-13377-C02 and No. FPA2011-29854-C04, the State of Hungary under Contract No. GINOP-2.3.3-15-2016-00034, the STFC (UK), the European Commission through the Marie Curie Actions Contract No. PIEFGA-2001-30096 and by the Japanese JSPS KAKENHI Grants No. 24740188 and No. 25247045.

\bibliographystyle{elsarticle-num}
\bibliography{bibliografia}

\begin{thebibliography}{10}
\expandafter\ifx\csname url\endcsname\relax
  \def\url#1{\texttt{#1}}\fi
\expandafter\ifx\csname urlprefix\endcsname\relax\def\urlprefix{URL }\fi
\expandafter\ifx\csname href\endcsname\relax
  \def\href#1#2{#2} \def\path#1{#1}\fi

\bibitem{Rac43}
G.~Racah, Phys. Rev. 63 (1943) 367.

\bibitem{Rac52}
G.~Racah, I.~Talmi, Physica 18 (1952) 1097.

\bibitem{Flo52}
B.~H. Flowers, Proceedings of the Royal Society of London A: Mathematical,
  Physical and Engineering Sciences 212~(1109) (1952) 248.

\bibitem{Talmi}
I.~Talmi, Simple Models of Complex Nuclei. The Shell Model and Interacting
  Boson Model, Harwood, Academic, Chur, Switzerland, 1993.

\bibitem{Casten}
R.~Casten, Nuclear Structure from a Simple Perspective, Oxford University
  press, New York, 2000.

\bibitem{Res04}
J.~J. Ressler, R.~F. Casten, N.~V. Zamfir, C.~W. Beausang, R.~B. Cakirli,
  H.~Ai, H.~Amro, M.~A. Caprio, A.~A. Hecht, A.~Heinz, S.~D. Langdown, E.~A.
  McCutchan, D.~A. Meyer, C.~Plettner, P.~H. Regan, M.~J.~S. Sciacchitano,
  A.~D. Yamamoto, Phys. Rev. C 69 (2004) 034317.

\bibitem{Esc06}
A.~Escuderos, L.~Zamick, Phys. Rev. C 73 (2006) 044302.

\bibitem{Gra02}
H.~Grawe, M.~G\`orska, C.~Fahlander, M.~Palacz, F.~Nowacki, E.~Caurier,
  J.~Daugas, M.~Lewitowicz, M.~Sawicka, R.~Grzywacz, K.~Rykaczewski, O.~Sorlin,
  S.~Leenhardt, F.~Azaiez, Nucl. Phys. A 704 (2002) 211.

\bibitem{Gor97}
M.~G\`orska, M.~Lipoglav\ifmmode~\check{s}\else \v{s}\fi{}ek, H.~Grawe,
  J.~Nyberg, A.~Atac̣, A.~Axelsson, R.~Bark, J.~Blomqvist, J.~Cederk\"all,
  B.~Cederwall, G.~de~Angelis, C.~Fahlander, A.~Johnson, S.~Leoni, A.~Likar,
  M.~Matiuzzi, S.~Mitarai, L.-O. Norlin, M.~Palacz, J.~Persson, H.~A. Roth,
  R.~Schubart, D.~Seweryniak, T.~Shizuma, O.~Skeppstedt, G.~Sletten, W.~B.
  Walters, M.~Weiszflog, Phys. Rev. Lett. 79 (1997) 2415.

\bibitem{Grz98}
R.~Grzywacz, R.~B\'eraud, C.~Borcea, A.~Emsallem, M.~Glogowski, H.~Grawe,
  D.~Guillemaud-Mueller, M.~Hjorth-Jensen, M.~Houry, M.~Lewitowicz, A.~C.
  Mueller, A.~Nowak, A.~P\l{}ochocki, M.~Pf\"utzner, K.~Rykaczewski, M.~G.
  Saint-Laurent, J.~E. Sauvestre, M.~Schaefer, O.~Sorlin, J.~Szerypo,
  W.~Trinder, S.~Viteritti, J.~Winfield, Phys. Rev. Lett. 81 (1998) 766.

\bibitem{Maz05}
C.~Mazzocchi, R.~Grzywacz, J.~Batchelder, C.~Bingham, D.~Fong, J.~Hamilton,
  J.~Hwang, M.~Karny, W.~Krolas, S.~Liddick, A.~Lisetskiy, A.~Morton,
  P.~Mantica, W.~Mueller, K.~Rykaczewski, M.~Steiner, A.~Stolz, J.~Winger,
  Phys. Lett. B 622 (2005) 45.

\bibitem{Jun07}
A.~Jungclaus, L.~C\'aceres, M.~G\`orska, M.~Pf\"utzner, S.~Pietri,
  E.~Werner-Malento, H.~Grawe, K.~Langanke, G.~Mart\'{\i}nez-Pinedo,
  F.~Nowacki, A.~Poves, J.~J. Cuenca-Garc\'{\i}a, D.~Rudolph, Z.~Podolyak,
  P.~H. Regan, P.~Detistov, S.~Lalkovski, V.~Modamio, J.~Walker, P.~Bednarczyk,
  P.~Doornenbal, H.~Geissel, J.~Gerl, J.~Grebosz, I.~Kojouharov, N.~Kurz,
  W.~Prokopowicz, H.~Schaffner, H.~J. Wollersheim, K.~Andgren, J.~Benlliure,
  G.~Benzoni, A.~M. Bruce, E.~Casarejos, B.~Cederwall, F.~C.~L. Crespi,
  B.~Hadinia, M.~Hellstr\"om, R.~Hoischen, G.~Ilie, J.~Jolie, A.~Khaplanov,
  M.~Kmiecik, R.~Kumar, A.~Maj, S.~Mandal, F.~Montes, S.~Myalski, G.~S.
  Simpson, S.~J. Steer, S.~Tashenov, O.~Wieland, Phys. Rev. Lett. 99 (2007)
  132501.

\bibitem{Got12}
A.~Gottardo, J.~J. Valiente-Dob\'on, G.~Benzoni, R.~Nicolini, A.~Gadea,
  S.~Lunardi, P.~Boutachkov, A.~M. Bruce, M.~G\'orska, J.~Grebosz, S.~Pietri,
  Z.~Podoly\'ak, M.~Pf\"utzner, P.~H. Regan, H.~Weick, J.~Alc\'antara
  N\'u\~nez, A.~Algora, N.~Al-Dahan, G.~de~Angelis, Y.~Ayyad, N.~Alkhomashi,
  P.~R.~P. Allegro, D.~Bazzacco, J.~Benlliure, M.~Bowry, A.~Bracco, M.~Bunce,
  F.~Camera, E.~Casarejos, M.~L. Cortes, F.~C.~L. Crespi, A.~Corsi, A.~M.
  Denis~Bacelar, A.~Y. Deo, C.~Domingo-Pardo, M.~Doncel, Z.~Dombradi,
  T.~Engert, K.~Eppinger, G.~F. Farrelly, F.~Farinon, E.~Farnea, H.~Geissel,
  J.~Gerl, N.~Goel, E.~Gregor, T.~Habermann, R.~Hoischen, R.~Janik, S.~Klupp,
  I.~Kojouharov, N.~Kurz, S.~M. Lenzi, S.~Leoni, S.~Mandal, R.~Menegazzo,
  D.~Mengoni, B.~Million, A.~I. Morales, D.~R. Napoli, F.~Naqvi, C.~Nociforo,
  A.~Prochazka, W.~Prokopowicz, F.~Recchia, R.~V. Ribas, M.~W. Reed,
  D.~Rudolph, E.~Sahin, H.~Schaffner, A.~Sharma, B.~Sitar, D.~Siwal,
  K.~Steiger, P.~Strmen, T.~P.~D. Swan, I.~Szarka, C.~A. Ur, P.~M. Walker,
  O.~Wieland, H.-J. Wollersheim, F.~Nowacki, E.~Maglione, A.~P. Zuker, Phys.
  Rev. Lett. 109 (2012) 162502.

\bibitem{Wat13}
H.~Watanabe, G.~Lorusso, S.~Nishimura, Z.~Y. Xu, T.~Sumikama, P.-A.
  S\"oderstr\"om, P.~Doornenbal, F.~Browne, G.~Gey, H.~S. Jung, J.~Taprogge,
  Z.~Vajta, J.~Wu, A.~Yagi, H.~Baba, G.~Benzoni, K.~Y. Chae, F.~C.~L. Crespi,
  N.~Fukuda, R.~Gernh\"auser, N.~Inabe, T.~Isobe, A.~Jungclaus, D.~Kameda,
  G.~D. Kim, Y.~K. Kim, I.~Kojouharov, F.~G. Kondev, T.~Kubo, N.~Kurz, Y.~K.
  Kwon, G.~J. Lane, Z.~Li, C.-B. Moon, A.~Montaner-Piz\'a, K.~Moschner,
  F.~Naqvi, M.~Niikura, H.~Nishibata, D.~Nishimura, A.~Odahara, R.~Orlandi,
  Z.~Patel, Z.~Podoly\'ak, H.~Sakurai, H.~Schaffner, G.~S. Simpson, K.~Steiger,
  H.~Suzuki, H.~Takeda, A.~Wendt, K.~Yoshinaga, Phys. Rev. Lett. 111 (2013)
  152501.

\bibitem{Sim14}
G.~S. Simpson, G.~Gey, A.~Jungclaus, J.~Taprogge, S.~Nishimura, K.~Sieja,
  P.~Doornenbal, G.~Lorusso, P.-A. S\"oderstr\"om, T.~Sumikama, Z.~Y. Xu,
  H.~Baba, F.~Browne, N.~Fukuda, N.~Inabe, T.~Isobe, H.~S. Jung, D.~Kameda,
  G.~D. Kim, Y.-K. Kim, I.~Kojouharov, T.~Kubo, N.~Kurz, Y.~K. Kwon, Z.~Li,
  H.~Sakurai, H.~Schaffner, Y.~Shimizu, H.~Suzuki, H.~Takeda, Z.~Vajta,
  H.~Watanabe, J.~Wu, A.~Yagi, K.~Yoshinaga, S.~B\"onig, J.-M. Daugas,
  F.~Drouet, R.~Gernh\"auser, S.~Ilieva, T.~Kr\"oll, A.~Montaner-Piz\'a,
  K.~Moschner, D.~M\"ucher, H.~Na\"{\i}dja, H.~Nishibata, F.~Nowacki,
  A.~Odahara, R.~Orlandi, K.~Steiger, A.~Wendt, Phys. Rev. Lett. 113 (2014)
  132502.

\bibitem{Qi11}
C.~Qi, Phys. Rev. C 83 (2011) 014307.

\bibitem{Zam07}
L.~Zamick, Phys. Rev. C 75 (2007) 064305.

\bibitem{Isa08}
P.~Van~Isacker, S.~Heinze, Phys. Rev. Lett. 100 (2008) 052501.

\bibitem{Qi10}
C.~Qi, Phys. Rev. C 81 (2010) 034318.

\bibitem{Isa13}
P.~Van~Isacker, I.~Celikovic, in: P.~Garret, B.~Hadinia (Eds.), Proc. 14th
  International Symposium on Capture Gamma-Ray Spectroscopy and Related Topics,
  World Scientific, 2013, p.~44.

\bibitem{Alh92}
Y.~Alhassid, A.~Leviatan, Journal of Physics A: Mathematical and General 25
  (1992) L1265.

\bibitem{Lev96}
A.~Leviatan, Phys. Rev. Lett. 77 (1996) 818.

\bibitem{Isa14}
P.~Van~Isacker, S.~Heinze, Annals of Physics 349~(Supplement C) (2014) 73.

\bibitem{Sod15}
P.-A. S\"oderstr\"om, S.~Nishimura, Z.~Y. Xu, K.~Sieja, V.~Werner,
  P.~Doornenbal, G.~Lorusso, F.~Browne, G.~Gey, H.~S. Jung, T.~Sumikama,
  J.~Taprogge, Z.~Vajta, H.~Watanabe, J.~Wu, H.~Baba, Z.~Dombradi, S.~Franchoo,
  T.~Isobe, P.~R. John, Y.-K. Kim, I.~Kojouharov, N.~Kurz, Y.~K. Kwon, Z.~Li,
  I.~Matea, K.~Matsui, G.~Mart\'{\i}nez-Pinedo, D.~Mengoni, P.~Morfouace, D.~R.
  Napoli, M.~Niikura, H.~Nishibata, A.~Odahara, K.~Ogawa, N.~Pietralla,
  E.~\ifmmode~\mbox{\c{S}}\else \c{S}\fi{}ahin, H.~Sakurai, H.~Schaffner,
  D.~Sohler, I.~G. Stefan, D.~Suzuki, R.~Taniuchi, A.~Yagi, K.~Yoshinaga, Phys.
  Rev. C 92 (2015) 051305.

\bibitem{Saw03}
M.~Sawicka, R.~Grzywacz, I.~Matea, H.~Grawe, M.~Pf\"utzner, J.~M. Daugas,
  M.~Lewitowicz, D.~L. Balabanski, F.~Becker, G.~B\'elier, C.~Bingham,
  C.~Borcea, E.~Bouchez, A.~Buta, M.~La~Commara, E.~Dragulescu, G.~de~France,
  G.~Georgiev, J.~Giovinazzo, M.~G\'orska, F.~Hammache, M.~Hass,
  M.~Hellstr\"om, F.~Ibrahim, Z.~Janas, H.~Mach, P.~Mayet, V.~M\'eot,
  F.~Negoita, G.~Neyens, F.~de~Oliveira~Santos, R.~D. Page, O.~Perru,
  Z.~Podoly\'ak, O.~Roig, K.~P. Rykaczewski, M.~G. Saint-Laurent, J.~E.
  Sauvestre, O.~Sorlin, M.~Stanoiu, I.~Stefan, C.~Stodel, C.~Theisen,
  D.~Verney, J.~\ifmmode~\dot{Z}\else \.{Z}\fi{}ylicz, Phys. Rev. C 68 (2003)
  044304.

\bibitem{Lis04}
A.~F. Lisetskiy, B.~A. Brown, M.~Horoi, H.~Grawe, Phys. Rev. C 70 (2004)
  044314.

\bibitem{Mor16}
A.~I. Morales, G.~Benzoni, H.~Watanabe, S.~Nishimura, F.~Browne, R.~Daido,
  P.~Doornenbal, Y.~Fang, G.~Lorusso, Z.~Patel, S.~Rice, L.~Sinclair, P.-A.
  S\"oderstr\"om, T.~Sumikama, J.~Wu, Z.~Y. Xu, A.~Yagi, R.~Yokoyama, H.~Baba,
  R.~Avigo, F.~L. Bello~Garrote, N.~Blasi, A.~Bracco, F.~Camera, S.~Ceruti,
  F.~C.~L. Crespi, G.~de~Angelis, M.-C. Delattre, Z.~Dombradi, A.~Gottardo,
  T.~Isobe, I.~Kojouharov, N.~Kurz, I.~Kuti, K.~Matsui, B.~Melon, D.~Mengoni,
  T.~Miyazaki, V.~Modamio-Hoyborg, S.~Momiyama, D.~R. Napoli, M.~Niikura,
  R.~Orlandi, H.~Sakurai, E.~Sahin, D.~Sohler, H.~Shaffner, R.~Taniuchi,
  J.~Taprogge, Z.~Vajta, J.~J. Valiente-Dob\'on, O.~Wieland, M.~Yalcinkaya,
  Phys. Rev. C 93 (2016) 034328.

\bibitem{Kol16}
K.~Kolos, D.~Miller, R.~Grzywacz, H.~Iwasaki, M.~Al-Shudifat, D.~Bazin, C.~R.
  Bingham, T.~Braunroth, G.~Cerizza, A.~Gade, A.~Lemasson, S.~N. Liddick,
  M.~Madurga, C.~Morse, M.~Portillo, M.~M. Rajabali, F.~Recchia, L.~L.
  Riedinger, P.~Voss, W.~B. Walters, D.~Weisshaar, K.~Whitmore, K.~Wimmer,
  J.~A. Tostevin, Phys. Rev. Lett. 116 (2016) 122502.

\bibitem{Mac17}
H.~Mach, A.~Korgul, M.~G\'orska, H.~Grawe, I.~Matea, M.~St\ifmmode~\u{a}\else
  \u{a}\fi{}noiu, L.~M. Fraile, Y.~E. Penionzkevich, F.~D.~O. Santos,
  D.~Verney, S.~Lukyanov, B.~Cederwall, A.~Covello, Z.~Dlouh\'y, B.~Fogelberg,
  G.~De~France, A.~Gargano, G.~Georgiev, R.~Grzywacz, A.~F. Lisetskiy,
  J.~Mrazek, F.~Nowacki, W.~A. P\l{}\'ociennik, Z.~Podoly\'ak, S.~Ray,
  E.~Ruchowska, M.-G. Saint-Laurent, M.~Sawicka, C.~Stodel, O.~Tarasov, Phys.
  Rev. C 95 (2017) 014313.

\bibitem{Qi17}
C.~Qi, Physics Letters B 773 (2017) 616.

\bibitem{Mar14}
T.~Marchi, G.~de~Angelis, J.~J. Valiente-Dob\'on, V.~M. Bader, T.~Baugher,
  D.~Bazin, J.~Berryman, A.~Bonaccorso, R.~Clark, L.~Coraggio, H.~L. Crawford,
  M.~Doncel, E.~Farnea, A.~Gade, A.~Gadea, A.~Gargano, T.~Glasmacher,
  A.~Gottardo, F.~Gramegna, N.~Itaco, P.~R. John, R.~Kumar, S.~M. Lenzi,
  S.~Lunardi, S.~McDaniel, C.~Michelagnoli, D.~Mengoni, V.~Modamio, D.~R.
  Napoli, B.~Quintana, A.~Ratkiewicz, F.~Recchia, E.~Sahin, R.~Stroberg,
  D.~Weisshaar, K.~Wimmer, R.~Winkler, Phys. Rev. Lett. 113 (2014) 182501.

\bibitem{Bro15}
F.~Browne, A.~Bruce, T.~Sumikama, I.~Nishizuka, S.~Nishimura, P.~Doornenbal,
  G.~Lorusso, P.-A. S\"oderstr\"om, H.~Watanabe, R.~Daido, Z.~Patel, S.~Rice,
  L.~Sinclair, J.~Wu, Z.~Y. Xu, A.~Yagi, H.~Baba, N.~Chiga, R.~Carroll,
  F.~Didierjean, Y.~Fang, N.~Fukuda, G.~Gey, E.~Ideguchi, N.~Inabe, T.~Isobe,
  D.~Kameda, I.~Kojouharov, N.~Kurz, T.~Kubo, S.~Lalkovski, Z.~Li, R.~Lozeva,
  H.~Nishibata, A.~Odahara, Z.~Podolyák, P.~H. Regan, O.~J. Roberts,
  H.~Sakurai, H.~Schaffner, G.~S. Simpson, H.~Suzuki, H.~Takeda, M.~Tanaka,
  J.~Taprogge, V.~Werner, O.~Wieland, Phys. Lett. B 750 (2015) 448.

\bibitem{Bro17}
F.~Browne, A.~M. Bruce, T.~Sumikama, I.~Nishizuka, S.~Nishimura, P.~Doornenbal,
  G.~Lorusso, P.-A. S\"oderstr\"om, H.~Watanabe, R.~Daido, Z.~Patel, S.~Rice,
  L.~Sinclair, J.~Wu, Z.~Y. Xu, A.~Yagi, H.~Baba, N.~Chiga, R.~Carroll,
  F.~Didierjean, Y.~Fang, N.~Fukuda, G.~Gey, E.~Ideguchi, N.~Inabe, T.~Isobe,
  D.~Kameda, I.~Kojouharov, N.~Kurz, T.~Kubo, S.~Lalkovski, Z.~Li, R.~Lozeva,
  N.~Nishibata, A.~Odahara, Z.~Podoly\'ak, P.~H. Regan, O.~J. Roberts,
  H.~Sakurai, H.~Schaffner, G.~S. Simpson, H.~Suzuki, H.~Takeda, M.~Tanaka,
  J.~Taprogge, V.~Werner, O.~Wieland, Phys. Rev. C 96 (2017) 024309.

\bibitem{Mor17}
A.~I. Morales, G.~Benzoni, H.~Watanabe, Y.~Tsunoda, T.~Otsuka, S.~Nishimura,
  F.~Browne, R.~Daido, P.~Doornenbal, Y.~Fang, G.~Lorusso, Z.~Patel, S.~Rice,
  L.~Sinclair, P.-A. S\"oderstr\"om, T.~Sumikama, J.~Wu, Z.~Y. Xu, A.~Yagi,
  R.~Yokoyama, H.~Baba, R.~Avigo, F.~L.~B. Garrote, N.~Blasi, A.~Bracco,
  F.~Camera, S.~Ceruti, F.~C.~L. Crespi, G.~de~Angelis, M.-C. Delattre,
  Z.~Dombradi, A.~Gottardo, T.~Isobe, I.~Kojouharov, N.~Kurz, I.~Kuti,
  K.~Matsui, B.~Melon, D.~Mengoni, T.~Miyazaki, V.~Modamio-Hoybjor,
  S.~Momiyama, D.~R. Napoli, M.~Niikura, R.~Orlandi, H.~Sakurai, E.~Sahin,
  D.~Sohler, H.~Schaffner, R.~Taniuchi, J.~Taprogge, Z.~Vajta, J.~J.
  Valiente-Dob\'on, O.~Wieland, M.~Yalcinkaya, Physics Letters B 765 (2017)
  328.

\bibitem{Tsu14}
Y.~Tsunoda, T.~Otsuka, N.~Shimizu, M.~Honma, Y.~Utsuno, Phys. Rev. C 89 (2014)
  031301.

\bibitem{Cor14}
L.~Coraggio, A.~Covello, A.~Gargano, N.~Itaco, Phys. Rev. C 89 (2014) 024319.

\bibitem{Leo17}
S.~Leoni, B.~Fornal, N.~M\ifmmode~\u{a}\else \u{a}\fi{}rginean, M.~Sferrazza,
  Y.~Tsunoda, T.~Otsuka, G.~Bocchi, F.~C.~L. Crespi, A.~Bracco, S.~Aydin,
  M.~Boromiza, D.~Bucurescu, N.~Cieplicka-Ory\ifmmode~\grave{n}\else
  \`{n}\fi{}czak, C.~Costache, S.~C\ifmmode~\u{a}\else \u{a}\fi{}linescu,
  N.~Florea, D.~G. Ghi\ifmmode \mbox{\c{t}}\else \c{t}\fi{}\ifmmode~\u{a}\else
  \u{a}\fi{}, T.~Glodariu, A.~Ionescu, L.~Iskra, M.~Krzysiek,
  R.~M\ifmmode~\u{a}\else \u{a}\fi{}rginean, C.~Mihai, R.~E. Mihai, A.~Mitu,
  A.~Negre\ifmmode~\mbox{\c{t}}\else \c{t}\fi{}, C.~R. Ni\ifmmode
  \mbox{\c{t}}\else \c{t}\fi{}\ifmmode~\u{a}\else \u{a}\fi{},
  A.~Ol\ifmmode~\u{a}\else \u{a}\fi{}cel, A.~Oprea, S.~Pascu, P.~Petkov,
  C.~Petrone, G.~Porzio, A.~\ifmmode~\mbox{\c{S}}\else \c{S}\fi{}erban,
  C.~Sotty, L.~Stan, I.~\ifmmode~\mbox{\c{S}}\else \c{S}\fi{}tiru, L.~Stroe,
  R.~\ifmmode \mbox{\c{S}}\else \c{S}\fi{}uv\ifmmode \u{a}\else
  \u{a}\fi{}il\ifmmode~\u{a}\else \u{a}\fi{}, S.~Toma,
  A.~Turturic\ifmmode~\u{a}\else \u{a}\fi{}, S.~Ujeniuc, C.~A. Ur, Phys. Rev.
  Lett. 118 (2017) 162502.

\bibitem{Sah17}
E.~Sahin, F.~L. Bello~Garrote, Y.~Tsunoda, T.~Otsuka, G.~de~Angelis,
  A.~G\"orgen, M.~Niikura, S.~Nishimura, Z.~Y. Xu, H.~Baba, F.~Browne, M.-C.
  Delattre, P.~Doornenbal, S.~Franchoo, G.~Gey, K.~Hady\ifmmode \acute{n}\else
  \'{n}\fi{}ska-Kl\ifmmode~\mbox{\c{e}}\else \c{e}\fi{}k, T.~Isobe, P.~R. John,
  H.~S. Jung, I.~Kojouharov, T.~Kubo, N.~Kurz, Z.~Li, G.~Lorusso, I.~Matea,
  K.~Matsui, D.~Mengoni, P.~Morfouace, D.~R. Napoli, F.~Naqvi, H.~Nishibata,
  A.~Odahara, H.~Sakurai, H.~Schaffner, P.-A. S\"oderstr\"om, D.~Sohler, I.~G.
  Stefan, T.~Sumikama, D.~Suzuki, R.~Taniuchi, J.~Taprogge, Z.~Vajta,
  H.~Watanabe, V.~Werner, J.~Wu, A.~Yagi, M.~Yalcinkaya, K.~Yoshinaga, Phys.
  Rev. Lett. 118 (2017) 242502.

\bibitem{Ben15}
G.~Benzoni, A.~I. Morales, H.~Watanabe, S.~Nishimura, L.~Coraggio, N.~Itaco,
  A.~Gargano, F.~Browne, R.~Daido, P.~Doornenbal, Y.~Fang, G.~Lorusso,
  Z.~Patel, S.~Rice, L.~Sinclair, P.-A. S\"oderstr\"om, T.~Sumikama, J.~Wu,
  Z.~Y. Xu, R.~Yokoyama, H.~Baba, R.~Avigo, F.~L.~B. Garrote, N.~Blasi,
  A.~Bracco, F.~Camera, S.~Ceruti, F.~C.~L. Crespi, G.~de~Angelis, M.-C.
  Delattre, Z.~Dombradi, A.~Gottardo, T.~Isobe, I.~Kuti, K.~Matsui, B.~Melon,
  D.~Mengoni, T.~Miyazaki, V.~Modamio-Hoybjor, S.~Momiyama, D.~R. Napoli,
  M.~Niikura, R.~Orlandi, H.~Sakurai, E.~Sahin, D.~Sohler, R.~Taniuchi,
  J.~Taprogge, Z.~Vajta, J.~J. Valiente-Dob\'on, O.~Wieland, M.~Yalcinkaya,
  Phys. Lett. B 751 (2015) 107.

\bibitem{Ots16}
T.~Otsuka, Y.~Tsunoda, J. Phys. G: Nucl. Part. Phys. 43 (2016) 024009.

\bibitem{Shi12}
N.~Shimizu, T.~Abe, Y.~Tsunoda, Y.~Utsuno, T.~Yoshida, T.~Mizusaki, M.~Honma,
  T.~Otsuka, Progress of Theoretical and Experimental Physics 2012 (2012)
  01A205.

\bibitem{Cor09}
L.~Coraggio, A.~Covello, A.~Gargano, N.~Itaco, T.~T.~S. Kuo, Progress in
  Particle and Nuclear Physics 62 (2009) 135.

\bibitem{Cor13}
L.~Coraggio, A.~Covello, A.~Gargano, N.~Itaco, T.~T.~S. Kuo, Annals of Physics
  327 (2012) 2125.

\bibitem{Wan17}
M.~Wang, G.~Audi, F.~Kondev, W.~Huang, S.~Naimi, X.~Xu, Chinese Physics C 41
  (2017) 030003.

\bibitem{Fuk13}
N.~Fukuda, T.~Kubo, T.~Ohnishi, N.~Inabe, H.~Takeda, D.~Kameda, H.~Suzuki,
  Nucl. Instrum. Methods B 317 (2013) 323.

\bibitem{ZeroDegree}
Y.~Mizoi, et~al., RIKEN Accel. Prog. Rep. 38 (2008) 297.

\bibitem{Nis12}
S.~Nishimura, Prog. Theor. Exp. Phys. 03C006.

\bibitem{Sod13}
P.-A. S\"odestr\"om, S.~Nishimura, P.~Doornenbal, G.~Lorusso, T.~Sumikama,
  H.~Watanabe, Z.~Xu, H.~Baba, F.~Browne, S.~Go, G.~Gey, T.~Isobe, H.-S. Jung,
  G.~Kim, Y.-K. Kim, I.~Kojouharov, N.~Kurz, Y.~Kwon, Z.~Li, K.~Moschner,
  T.~Nakao, H.~Nishibata, M.~Nishimura, A.~Odahara, H.~Sakurai, H.~Schaffner,
  T.~Shimoda, J.~Taprogge, Z.~Vajta, V.~Werner, J.~Wu, A.~Yagi, K.~Yoshinaga,
  Nucl. Instrum. Methods B 317 (2012) 649.

\bibitem{Mor15_proc}
A.~I. Morales, G.~Benzoni, H.~Watanabe, D.~Sohler, E.~Sahin, G.~de~Angelis,
  S.~Nishimura, G.~Lorusso, T.~Sumikama, P.~Doornenbal, Z.~Xu, T.~Isobe,
  P.~S\"oderstr\"om, F.~Browne, J.~Wu, H.~Baba, Z.~Patel, S.~Rice, L.~Sinclair,
  R.~Yokoyama, R.~Daido, Y.~Fang, M.~Niikura, R.~Avigo, F.~B. Garrote,
  N.~Blasi, A.~Bracco, S.~Ceruti, F.~Crespi, M.-C. Delattre, Z.~Dombradi,
  A.~Gottardo, I.~Kuti, K.~Matsui, B.~Melon, D.~Mengoni, T.~Miyazaki,
  V.~Modamio-Hoybjor, S.~Momiyama, D.~Napoli, R.~Orlandi, H.~Sakurai,
  R.~Taniuchi, J.~Taprogge, Z.~Vajta, J.~Valiente-Dob\'on, O.~Wieland, A.~Yagi,
  M.~Yalcinkaya, in: Exotic Nuclei, World Scientific, 2015, Ch.~45, p. 429.

\bibitem{Pat14}
Z.~Patel, et~al., RIKEN Acc. Prog. Rep. 47 (2014) 13.

\bibitem{Mor17b}
A.~I. Morales, A.~Algora, B.~Rubio, K.~Kaneko, S.~Nishimura, P.~Aguilera,
  S.~E.~A. Orrigo, F.~Molina, G.~de~Angelis, F.~Recchia, G.~Kiss, V.~H. Phong,
  J.~Wu, D.~Nishimura, H.~Oikawa, T.~Goigoux, J.~Giovinazzo, P.~Ascher,
  J.~Agramunt, D.~S. Ahn, H.~Baba, B.~Blank, C.~Borcea, A.~Boso, P.~Davies,
  F.~Diel, Z.~Dombr\'adi, P.~Doornenbal, J.~Eberth, G.~de~France, Y.~Fujita,
  N.~Fukuda, E.~Ganioglu, W.~Gelletly, M.~Gerbaux, S.~Gr\'evy, V.~Guadilla,
  N.~Inabe, T.~Isobe, I.~Kojouharov, W.~Korten, T.~Kubo, S.~Kubono,
  T.~Kurtuki\'an~Nieto, N.~Kurz, J.~Lee, S.~Lenzi, J.~Liu, T.~Lokotko,
  D.~Lubos, C.~Magron, A.~Montaner-Piz\'a, D.~R. Napoli, H.~Sakurai,
  H.~Schaffner, Y.~Shimizu, C.~Sidong, P.-A. S\"oderstr\"om, T.~Sumikama,
  H.~Suzuki, H.~Takeda, Y.~Takei, M.~Tanaka, S.~Yagi, Phys. Rev. C 95 (2017)
  064327.

\bibitem{Paa79}
V.~Paar, Nucl. Phys. A 331 (1979) 16.

\bibitem{Van85}
J.~Van~Maldeghem, K.~Heyde, J.~Sau, Phys. Rev. C 32 (1985) 1067.

\bibitem{Mol00}
W.~F. Mueller, B.~Bruyneel, S.~Franchoo, M.~Huyse, J.~Kurpeta, K.~Kruglov,
  Y.~Kudryavtsev, N.~V. S.~V. Prasad, R.~Raabe, I.~Reusen, P.~Van~Duppen,
  J.~Van~Roosbroeck, L.~Vermeeren, L.~Weissman, Z.~Janas, M.~Karny, T.~Kszczot,
  A.~P\l{}ochocki, K.-L. Kratz, B.~Pfeiffer, H.~Grawe, U.~K\"oster, P.~Thirolf,
  W.~B. Walters, Phys. Rev. C 61 (2000) 054308.

\bibitem{spoiler}
C.~J. Prokop, B.~P. Crider, S.~N. Liddick, A.~D. Ayangeakaa, M.~P. Carpenter,
  J.~J. Carroll, J.~Chen, C.~J. Chiara, H.~M. David, A.~C. Dombos, S.~Go,
  J.~Harker, R.~V.~F. Janssens, N.~Larson, T.~Lauritsen, R.~Lewis, S.~J. Quinn,
  F.~Recchia, D.~Seweryniak, A.~Spyrou, S.~Suchyta, W.~B. Walters, S.~Zhu,
  Phys. Rev. C 92 (2015) 061302.

\bibitem{Cri16}
B.~P. Crider, C.~J. Prokop, S.~Liddick, M.~Al-Shudifat, A.~D. Ayangeakaa, M.~P.
  Carpenter, J.~J. Carroll, J.~Chen, C.~J. Chiara, H.~M. David, A.~C. Dombos,
  S.~Go, R.~Grzywacz, J.~Harker, R.~V.~F. Janssens, N.~Larson, T.~Lauritsen,
  R.~Lewis, S.~J. Quinn, F.~Recchia, A.~Spyrou, S.~Suchyta, W.~B. Walters,
  S.~Zhu, Physics Letters B 763~(Supplement C) (2016) 108.

\end{thebibliography}

\end{document}